\documentclass[preprint]{aastex}
\usepackage{emulateapj5}

\newcommand{\Msun}{$M_{\odot}$}
\submitted{Accepted for publication in the Astrophysical Journal}

\shortauthors{METCALFE, NATHER, \& WINGET}
\shorttitle{GENETIC-ALGORITHM-BASED ANALYSIS OF GD~358}

\begin{document}

\title{Genetic-Algorithm-based Asteroseismological Analysis\\
        of the DBV White Dwarf GD~358}

\author{T.S. Metcalfe, R.E. Nather, and D.E. Winget} 
\affil{Department of Astronomy, University of Texas at Austin}
\affil{Mail Code C1400, Austin, TX 78712}
\email{travis@astro.as.utexas.edu}

\begin{abstract}

White dwarf asteroseismology offers the opportunity to probe the structure
and composition of stellar objects governed by relatively simple physical
principles. The observational requirements of asteroseismology have been
addressed by the development of the Whole Earth Telescope (WET), but the
analytical procedures still need to be refined before this technique can
yield the complete physical insight that the data can provide. Toward this
end, we have utilized a genetic-algorithm-based optimization method to fit
our models to the observed pulsation frequencies of the DBV white dwarf
GD~358 obtained with the WET in 1990. This new approach has finally
exploited the sensitivity of our models to the core composition, and will
soon yield some interesting constraints on nuclear reaction
cross-sections. 

\end{abstract}


\section{INTRODUCTION}

We study white dwarf stars because they are the end-points of stellar
evolution for the majority of all stars, and their composition and
structure can tell us about their prior history. We can determine the
internal structure of pulsating white dwarfs by observing their variations
in brightness over time, using the techniques of high speed photometry to
define their light curves, and then matching these observations with a
numerical model which behaves the same way. The parameters of the model
are chosen to correspond one-to-one with the physical processes that give
rise to the variations, so a good fit to the data leads us to believe that
our model reflects the actual physics of the stars themselves. 

Although this procedure is simple in outline, its realization in practice
requires specialized instrumentation to overcome the practical
difficulties we encounter in the process. The Whole Earth Telescope (WET) 
observing network \citep{nat90} was developed to provide the 200 or more
hours of essentially gap-free data we require for the analysis. This
instrument is now mature, and has provided a wealth of seismological data
on the different varieties of pulsating white dwarf stars, so the
observational part of the procedure is well in hand. Now we are working to
improve our analytical procedures to take full advantage of the
possibilities afforded by asteroseismology. 

The adjustable parameters in our computer models of white dwarfs presently
include the total mass, the temperature, hydrogen and helium layer masses,
core composition, convective efficiency, and composition transition
profiles. Finding a proper set of these to provide a close fit to the
observed data is difficult. The traditional procedure is a cut-and-try
process guided by intuition and experience, and is far more subjective
than we would like. More objective procedures are essential if
asteroseismology is to become a widely-accepted astronomical technique. We
must be able to demonstrate that, within the range of different values the
model parameters can assume, we have found the only solution or the best
one if more than one is possible.

\section{GENETIC ALGORITHMS}

An optimization scheme based on a genetic algorithm (GA) can avoid the
problems inherent in the traditional approach. Restrictions on the range
of the parameter-space are imposed only by observational constraints and
by the physics of the model. Although the parameter-space so defined is
often quite large, the GA provides a relatively efficient means of
searching globally for the best-fit model. While it is difficult for GAs
to find precise values for the best-fit set of parameters, they are very
good at finding the region of parameter-space that contains the global
minimum. In this sense, the GA is an objective means of finding a good
first guess for a more traditional method which can then narrow in on the
precise values and uncertainties of the best-fit set of parameters. 

The underlying ideas for genetic algorithms were inspired by Charles
Darwin's \citeyearpar{dar59} notion of biological evolution through
natural selection. The basic idea is to solve an optimization problem by
{\it evolving} the best solution from an initial set of completely random
guesses. The theoretical model provides the framework within which the
evolution takes place, and the individual parameters controlling it serve
as the genetic building blocks. Observations provide the selection
pressure. For a detailed description of genetic algorithms, see
\cite{met99} and \cite{cha95}. 

Initially, the parameter-space is filled uniformly with trials consisting
of randomly chosen values for each parameter, within a range based on the
physics that the parameter is supposed to describe. The model is evaluated
for each trial, and the result is compared to the observed data and
assigned a {\it fitness} inversely proportional to the variance. A new
generation of trials is then created by selecting from this population at
random, weighted by the fitness. 

Each model is encoded into a long string of numbers analogous to a
chromosome with each parameter serving as a gene. The encoded trials are
paired up and modified in order to explore new regions of parameter-space. 
The two basic genetic operators are {\it crossover} which emulates
reproduction, and {\it mutation} which emulates happenstance. After these
operators have been applied, the strings are decoded back into sets of
numerical values for the parameters. The new generation replaces the old
one, and the process begins again. The evolution continues for a specified
number of generations, chosen to maximize the efficiency of the method.

\section{THE DBV WHITE DWARF GD 358}

During a survey of eighty-six suspected white dwarf stars in the Lowell GD
lists, \cite{gre69} classified GD 358 as a helium atmosphere (DB) white
dwarf based on its spectrum. Photometric $UBV$ and $ubvy$ colors were
later determined by \cite{bw73} and \cite{weg79} respectively. Time-series
photometry by \cite{wrn82} revealed the star to be a pulsating
variable---the first confirmation of a new class of variable (DBV) white
dwarfs predicted by \cite{win81}. 

In May 1990, GD 358 was the target of a coordinated observing run with the
WET. The results of these observations were reported by \cite{win94}, and
the theoretical interpretation was given in a companion paper by
\cite{bw94b}. They found a series of nearly equally-spaced periods in the
power spectrum which they interpreted as non-radial {\it g}-mode
pulsations of consecutive radial overtone.  They attempted to match the
observed periods and the period spacing for these modes using earlier
versions of the same theoretical models we have used in this analysis (see
\S \ref{MODSECT}). Their optimization method involved computing a grid of
models near a first guess determined from general scaling arguments and
analytical relations developed by \cite{kaw90}, \cite{kw90},
\cite{bfwh92}, and \cite{bww93}.

\section{DBV WHITE DWARF MODELS \label{modsect}}

\subsection{Defining the Parameter-Space}

The most important parameters affecting the pulsation properties of DBV
white dwarf models are the total stellar mass ($M_*$), the effective
temperature ($T_{\rm eff}$), and the mass of the atmospheric helium layer
($M_{\rm He}$). We want to be careful to avoid introducing any subjective
bias into the best-fit determination simply by defining the range of the
search too narrowly. For this reason, we have specified the range for each
parameter based only on the physics of the model, and on observational
constraints.

The distribution of masses for isolated white dwarf stars, generally
inferred from measurements of $\log\ g$, is strongly peaked near 0.6
\Msun\ with a FWHM of about 0.1 \Msun\ \citep{ngs99}. Isolated main
sequence stars with masses near the limit for helium ignition produce C/O
cores more massive than about 0.45 \Msun, so white dwarfs with masses
below this limit must have helium cores \citep{sgr90,ngs99}. However, the
universe is not presently old enough to produce helium core white dwarfs
through single star evolution.  We confine our search to masses between
0.45 \Msun\ and 0.95 \Msun. Although some white dwarfs
 
\epsfxsize 3.5in
\epsffile{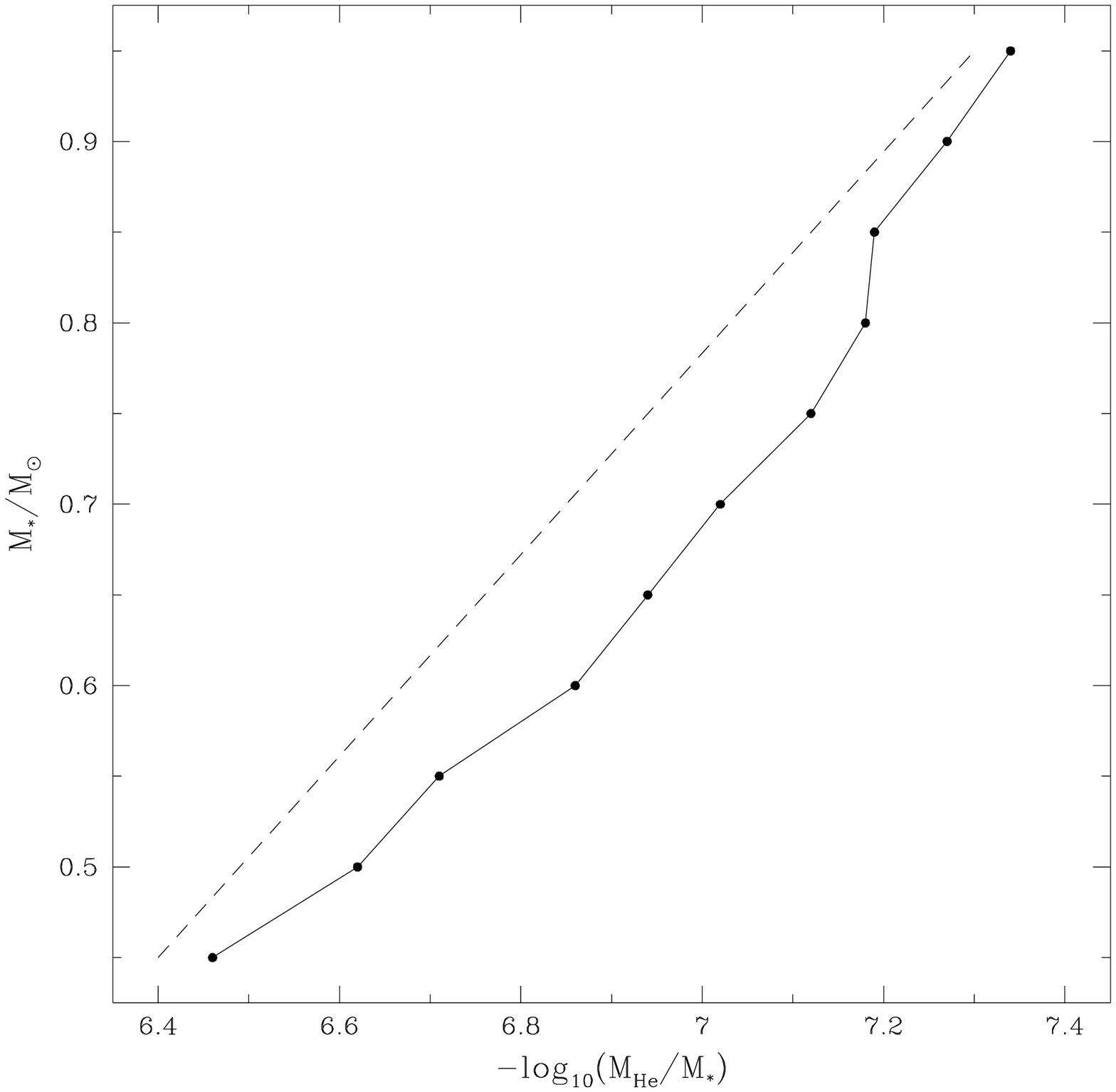}
\figcaption[gd358.fig1.ps]{The numerical thin limit of the fractional 
helium layer mass for various values of the total mass (connected points) 
and the linear cut implemented in the WD-40 code (dashed line). 
\label{fig1}} 
\vskip 18pt

\noindent are known to be
more massive than the upper limit of our search, these represent a very
small fraction of the total population, and for reasonable assumptions
about the mass-radius relation all known DBVs appear to have masses within
the range of our search \citep{bea99}. 

The span of effective temperatures within which DB white dwarfs are
pulsationally unstable is known as the DB instability strip.  The precise
location of this strip is the subject of some debate, primarily because of
difficulties in matching the temperature scales from ultraviolet and
optical spectroscopy and the possibility of hiding trace amounts of
hydrogen in the envelope \citep{bea99}. The most recent temperature
determinations for the 8 known DBV stars were done by \cite{bea99}. These
measurements, depending on various assumptions, place the red edge as low
as 21,800 K, and the blue edge as high as 27,800 K. Our search includes
all temperatures between 20,000 K and 30,000 K.

The mass of the atmospheric helium layer must not be greater than about
$10^{-2}\ M_*$ or the pressure of the overlying material would
theoretically initiate helium burning at the base of the envelope. At the
other extreme, none of our models pulsate for helium layer masses less
than about $10^{-8}\ M_*$ over the entire temperature range we are
considering \citep{bw94a}. The practical limit is actually slightly larger
than this theoretical limit, and is a function of mass. For the most
massive white dwarfs we consider, our models run smoothly with a helium
layer as thin as $5\times 10^{-8}\ M_*$, while for the least massive the
limit is $4\times 10^{-7}\ M_*$ (see Figure 1). 

\subsection{Theoretical Models}

To find the theoretical pulsation modes of a white dwarf, we start with a
static model of a pre-white dwarf and allow it to evolve quasi-statically
until it reaches the desired temperature. We then calculate the adiabatic
non-radial oscillation frequencies for the output model.  The initial
`starter' models can come from detailed calculations that evolve a
main-sequence star all the way to its pre-white dwarf phase, but this is
generally only important for accurate models of the hot DO white dwarfs. 
For the cooler DB and DA white dwarfs, it is sufficient to start with a
hot polytrope of order 2/3 (i.e. $P\propto\rho^{5/3}$).  The cooling
tracks of these polytropes converge with those of the pre-white dwarf
models well above the temperatures at which DB and DA white dwarfs are
observed to be pulsationally unstable \citep{woo90}.

To allow fitting for the total mass, we generated a grid of 100 starter
models with masses between 0.45 and 0.95 \Msun. The entire grid originated
from a 0.6 \Msun\ carbon-core polytrope starter model. We performed a
homology transform on this model to generate three new masses: 0.65, 0.75,
and 0.85 \Msun. We relaxed each of these three, and then used all four to
generate the grid. All models with masses below 0.6 \Msun\ were generated
by a direct homology transform of the original 0.6 \Msun\ polytrope. For
masses between $0.605 \rightarrow 0.745$ \Msun\ and from $0.755
\rightarrow 0.895$ \Msun, we used homology transforms of the relaxed 0.65
\Msun\ and 0.75 \Msun\ models respectively. The models with masses greater
than 0.9 \Msun\ were homology transforms of the relaxed 0.85 \Msun\ model.

To evolve a starter model to a specific temperature, we used the White
Dwarf Evolution Code (WDEC) described in detail by \cite{lv75} and by
\cite{woo90}. This code was originally written by Martin Schwarzschild,
and has subsequently been updated and modified by many others including:
\cite{ks69}, \cite{lv75}, \cite{win81}, \cite{kaw86}, \cite{woo90},
\cite{bra93}, and \cite{mon98}. The equation of state (EOS) for the cores
of our models come from \cite{lam74}, and from \cite{fgv77} for the
envelopes. We use the updated OPAL opacity tables from \cite{ir93},
neutrino rates from \cite{ito89}, and the ML3 mixing-length prescription
of \cite{bc71}. The evolution calculations for the core are fully
self-consistent, but the envelope is treated separately. The core and
envelope are stitched together and the envelope is adjusted to match the
boundary conditions at the interface. Adjusting the helium layer mass
involves stitching an envelope with the desired thickness onto the core
before starting the evolution. Because this is done while the model is
still very hot, there is plenty of time to reach equilibrium before the
model approaches the final temperature. 

We determined the pulsation frequencies of the output models using the
adiabatic non-radial oscillation (ANRO) code described by \cite{kaw86},
originally written by Carl Hansen, which solves the pulsation equations
using the Runge-Kutta-Fehlberg method.

We have made extensive practical modifications to these programs,
primarily to allow models to be calculated without any intervention by the
user. The result is a combined evolution/pulsation code that runs smoothly
over a wide range of input parameters. We call this new code WD-40. Given
a mass, temperature, and helium layer mass within the ranges discussed
above, WD-40 will evolve and pulsate the specified white dwarf model and
return a list of the theoretical pulsation periods.

\section{MODEL FITTING}

The execution time for a single DB model on a reasonably fast computer is
less than a minute. The genetic algorithm approach, however, requires the
evaluation of many thousands of such models. To be practical, we ran the
code in parallel on a specialized computational instrument---a collection
of minimal PCs connected by a network---which we designed and built for
this project \citep[see][]{mn99a,mn99b}.

We used the most recent version of a public-domain GA (called PIKAIA) 
described in detail by \cite{cha95}. To allow the evaluation of models in
parallel, we incorporated the message passing routines of the Parallel
Virtual Machine (PVM) software \citep{gei94} into the ``full generational
replacement'' evolution option of PIKAIA. 

Our strategy was to use PIKAIA as a {\it master} program to exchange data
with many copies of the WD-40 code serving as {\it slave} processes.  The
master program sends sets of parameters to slave processes running on
every available processor, and then listens for responses. It sends new
jobs after each response, and continues until all of the trials for a
particular generation have been calculated. The master program then
performs the genetic shuffling to come up with a new generation of trials
to be calculated. This continues for a specified number of generations,
and the best solution in the final population of trials is used as the
first guess for a more traditional approach.

The slave processes, when they receive data from the master program,
evolve a white dwarf model with the specified properties, calculate the
pulsation periods, and compare them to the observed periods. The relative
fitness of the trial---which we define as the inverse of the
root-mean-square (r.m.s.) residuals between the observed and calculated
pulsation periods---is then sent back to the master program. 

We used a population size of 128 trials, and initially allowed the GA to
run for 250 generations. We used 2-digit decimal encoding for each of the
three parameters, which resulted in a temperature resolution of 100 K, a
mass resolution of 0.005 \Msun, and a resolution for the helium layer
thickness of 0.05 dex. The uniform single-point crossover probability was
fixed at 85\%, and the mutation rate was allowed to vary between 0.1\% and
16.6\%, depending on the linear distance in parameter-space between the
trials with the median and the best fitnesses.

\subsection{Application to Noiseless Simulated Data \label{inmodsect}}

To quantify the efficiency of our method for this problem, we used the
WD-40 code to calculate the pulsation periods of a model within the search
space, and then attempted to find the set of input parameters [$T_{\rm
eff}=25,000$~K, $M_* = 0.600$ \Msun, $\log (M_{\rm He}/M_*) = -5.96$]
using the GA. We performed 20 independent runs using different
pseudo-random number sequences each time. The first order solutions found
in each case by the GA are listed in Table 1.  In 9 of the 20 runs, the GA
found the exact set of input parameters, and in 4 other runs it finished
in a region of parameter-space close enough for a small (1331 point) grid
to reveal the exact answer. Since none

\begin{table*}
\begin{center}
\tablenum{1}\label{tab1}
\centerline{\sc Table 1}
\centerline{\sc Results for Noiseless Simulated Data}\vskip 6pt
\begin{tabular}{cccccc}
\hline\hline
& \multicolumn{3}{c}{First-Order Solution} & & Generation \\ \cline{2-4}
Run & $T_{\rm eff}$ & $M/M_{\odot}$ & $\log(M_{\rm He}/M_*)$ & r.m.s.& Found \\
\hline
01 & 26,800 & 0.560 & $-$5.70 & 0.67 & 245 \\
02 & 25,000 & 0.600 & $-$5.96 & 0.00 & 159 \\
03 & 24,800 & 0.605 & $-$5.96 & 0.52 & 145 \\
04 & 25,000 & 0.600 & $-$5.96 & 0.00 &  68 \\
05 & 22,500 & 0.660 & $-$6.33 & 1.11 &  97 \\
06 & 25,000 & 0.600 & $-$5.96 & 0.00 & 142 \\
07 & 25,000 & 0.600 & $-$5.96 & 0.00 &  97 \\
08 & 25,000 & 0.600 & $-$5.96 & 0.00 & 194 \\
09 & 25,200 & 0.595 & $-$5.91 & 0.42 & 116 \\
10 & 26,100 & 0.575 & $-$5.80 & 0.54 &  87 \\
11 & 23,900 & 0.625 & $-$6.12 & 0.79 &  79 \\
12 & 25,000 & 0.600 & $-$5.96 & 0.00 & 165 \\
13 & 26,100 & 0.575 & $-$5.80 & 0.54 &  92 \\
14 & 25,000 & 0.600 & $-$5.96 & 0.00 &  95 \\
15 & 24,800 & 0.605 & $-$5.96 & 0.52 &  42 \\
16 & 26,600 & 0.565 & $-$5.70 & 0.72 & 246 \\
17 & 24,800 & 0.605 & $-$5.96 & 0.52 & 180 \\
18 & 25,000 & 0.600 & $-$5.96 & 0.00 &  62 \\
19 & 24,100 & 0.620 & $-$6.07 & 0.76 & 228 \\
20 & 25,000 & 0.600 & $-$5.96 & 0.00 & 167 \\
\hline\hline
\end{tabular}
\end{center}
\end{table*}

\noindent of the successful runs converged
between generations 200 and 250, we stopped future runs after 200
generations.

From the 13 runs that converged in 200 generations, we deduce an
efficiency for the method (GA + small grid) of $\sim$65\%. This implies
that the probability of missing the correct answer in a single run is
$\sim$35\%.  By running the GA several times, we reduce the probability of
not finding the correct answer: the probability that two runs will both be
wrong is $\sim$12\%, for three runs it is $\sim$4\%, and so on. Thus, to
reduce the probability of not finding the correct answer to below 1\% we
need to run the GA, on average, 5 times. For 200 generations of 128
trials, this requires $\sim$10$^5$ model evaluations. By comparison, an
exhaustive search of the parameter-space with the same resolution would
require $10^6$ model evaluations, so our method is comparably global but
$\sim$10$\times$ more efficient than an exhaustive search of
parameter-space. Even with this efficiency and our ability to run the
models in parallel, each run of the GA required about 6 hours to complete. 

\subsection{The Effect of Gaussian Noise \label{noisesect}}

Having established that the GA could find the correct answer for noiseless
data, we wanted to see how noise on the frequencies might affect it.
Before adding noise to the input frequencies, we attempted to characterize
the actual noise present on frequencies determined from a WET campaign. We
approached this problem in two different ways.

First, we tried to characterize the noise empirically using the mode
identifications from \cite{vui00} to look at the distribution of
differences between the observed and predicted linear combination
frequencies in the 1994 WET run on GD 358. There were a total of 63
combinations identified: 20 sum and 11 difference frequencies of 2-mode
combinations, 30 sum and difference 3-mode combinations, and 2
combinations involving 4 modes. We used the measured frequencies of the
parent modes to predict the frequency of each linear combination, and then
compared this to the observed frequency.  The distribution of observed
minus computed frequencies for these 63 modes, and the best-fit Gaussian
is shown in the top panel of Figure \ref{fig2}. The distribution has
$\sigma=0.17\ \mu$Hz.

Second, we tried to characterize the noise by performing the standard
analysis for WET runs on many simulated light curves to look at the
distribution of differences between the input and output frequencies. We
generated 100 synthetic GD 358 light curves using the 57 observed
frequencies and amplitudes from \cite{win94}. Each light curve had the
same time span as the 1990 WET run (965,060 seconds) sampled with the same
interval (every 10 seconds) but without any gaps in coverage. Although the
noise in the observed light curves was clearly time-correlated, we found
that the distribution around the mean light level for the comparison star
after the standard reduction procedure was well represented by a Gaussian. 
So we added Gaussian noise to the simulated light curves to yield a
signal-to-noise ratio $S/N \approx 2$, which is typical of the observed
data. We took the discrete Fourier Transform of each light curve, and
identified peaks in the same way as is done for real WET runs. We
calculated the differences between the input frequencies and those
determined from the simulation for the 11 modes used in the seismological
analysis by \cite{bw94b}. The distribution of these differences is shown
in the bottom panel of Figure \ref{fig2}, along with the best-fit Gaussian
which has $\sigma=0.053\ \mu$Hz. We adopted this value for our studies of
the effects of noise on the GA method.

Using the same input model as in \S \ref{inmodsect}, we added random
offsets drawn from a Gaussian distribution with $\sigma=0.053\ \mu$Hz to
each of the frequencies.  We produced 10 sets 

\epsfxsize 3.5in
\epsffile{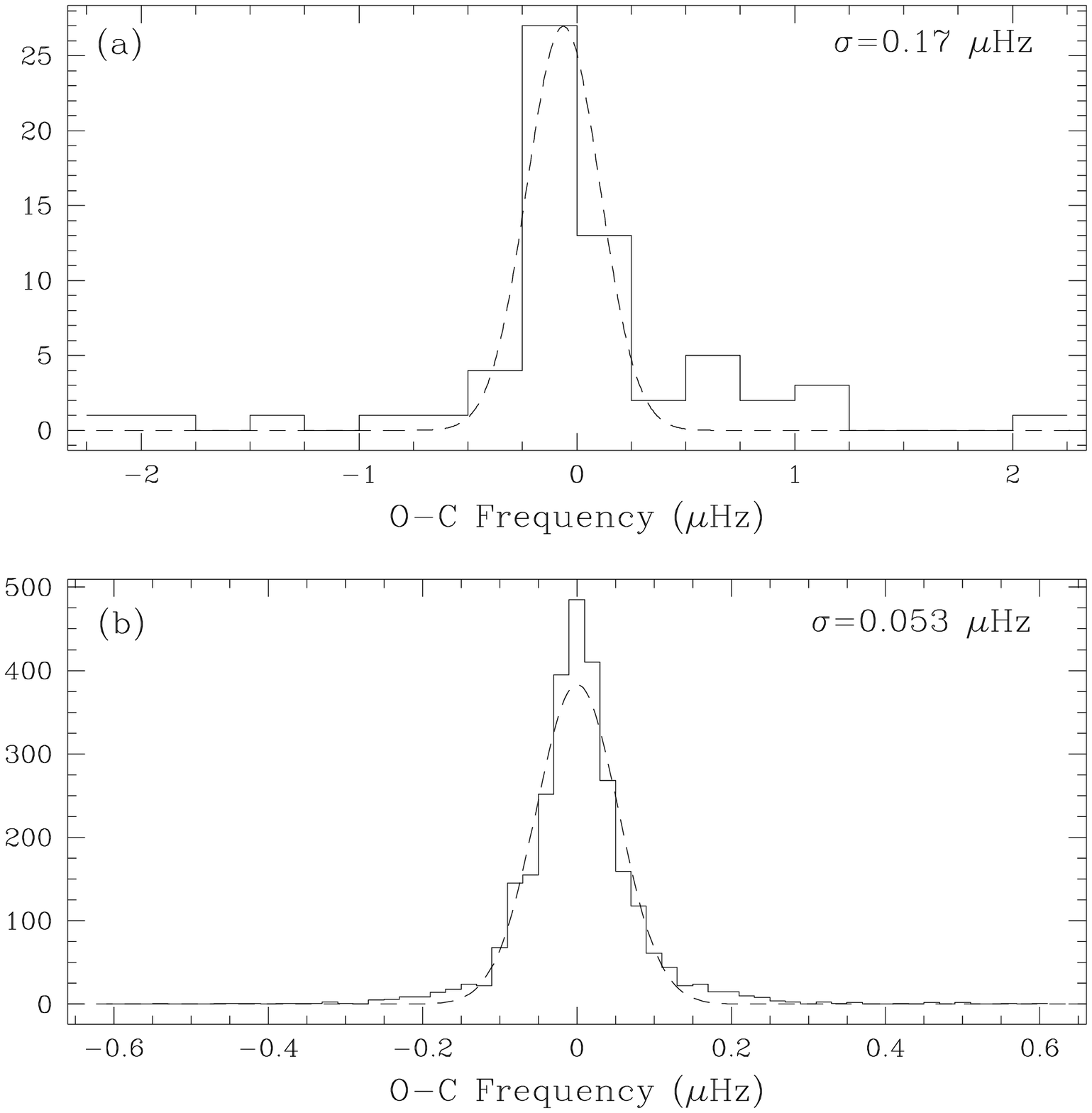}
\figcaption[gd358.fig2.ps]{The distribution of differences between (a) the 
observed and predicted frequencies of linear combination modes identified
by \cite{vui00} in the 1994 WET run on GD~358 and the best-fit Gaussian 
with $\sigma=0.17\ \mu$Hz (dashed line) and (b) the input and output 
frequencies of the 11 modes used to model GD~358 from simulated WET runs 
(see \S \ref{noisesect} for details) and the best-fit Gaussian with 
$\sigma=0.053\ \mu$Hz (dashed line).\label{fig2}}
\vskip 10pt

\noindent of target frequencies from
10 unique realizations of the noise, and then applied the GA method to
each set.  In all cases the best of 5 runs of the GA found the exact set
of parameters from the original input model, or a region close enough to
reveal the exact solution after calculating the small grid around the
first guess.  To reassure ourselves that the success of the GA did not
depend strongly on the amount of noise added to the frequencies, we also
applied the GA to several realizations of the larger noise estimate from
the analysis of linear combination frequencies. The GA always succeeded in
finding the original input model.

\subsection{Application to GD 358}

Having thoroughly characterized the GA method, we finally applied it to
real data. We used the same 11 periods used by \cite{bw94b}. As in their
analysis, we assumed that the periods were consecutive radial overtones
and that they were all $\ell=1$ modes. Anticipating that the GA might have
more difficulty with non-synthetic data, we decided to perform a total of
10 GA runs for each core composition. This should reduce the chances of
not finding the best answer to less than about 3 in 10,000. 

To facilitate comparison with previous results, we obtained fits for six
different combinations of core composition and transition profiles: pure
C, pure O, and both ``steep'' and ``shallow'' transition profiles for
50:50 and 20:80 C/O cores \citep[see][]{bww93,bw94b}.

We also ran the GA with an alternate fitness criterion for the 50:50 C/O
``steep'' case, which contains the best-fit model of \cite{bw94b}.
Normally, the GA only attempts to match the pulsation periods. We
reprogrammed it to match both the periods and deviations from the mean
period spacing, which was the fitness criterion used by
\citeauthor{bw94b}. Within the range of parameters they considered, using
this alternate fitness criterion, the GA found best-fit model parameters
consistent with \citeauthor{bw94b}'s solution.

\section{RESULTS}

\begin{figure*}
\epsfxsize 7.0in
\epsffile{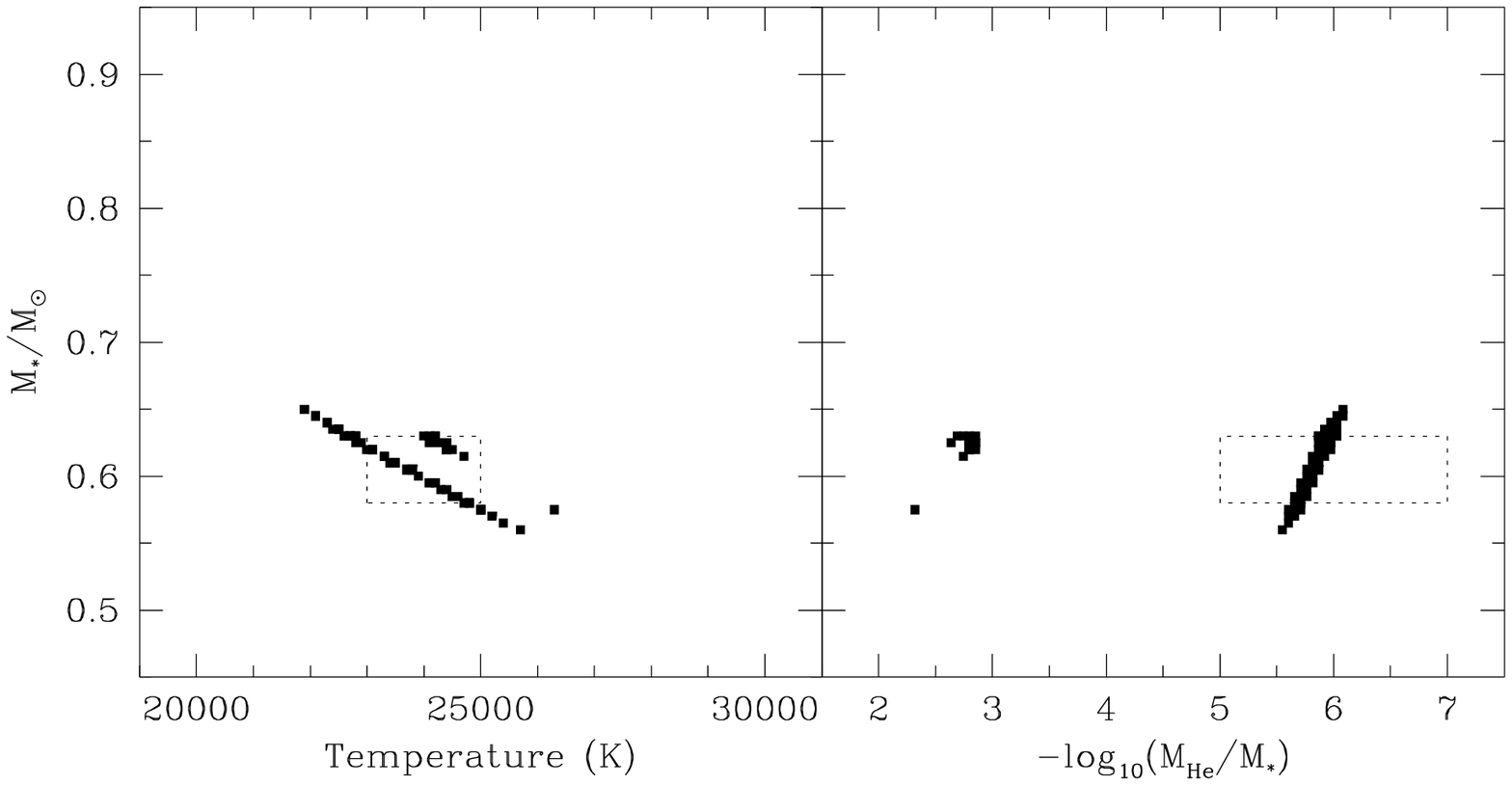}
\figcaption[gd358.fig3.ps]{Front and side views of the GA search space for
a C/O 50:50 core with a ``steep'' transition profile. Square points mark the
locations of every model found by the GA with an r.m.s. deviation smaller 
than 3 seconds for the periods observed in GD 358. The dashed line shows
the range of parameters considered by \cite{bw94b}.
\label{fig3}}
\end{figure*}

The general features of the 3-dimensional parameter-space [$M_*, T_{\rm
eff}, -\log_{10}(M_{\rm He}/M_*)$] for GD~358 are illustrated in Figure
\ref{fig3}.  All combinations of parameters found by the GA for a 50:50
C/O steep core having r.m.s.~period differences smaller than 3 seconds are
shown as square points in this plot. The two panels are orthogonal
projections of the search space, so each point in the left panel
corresponds one-to-one with a point in the right panel. Essentially,
Figure \ref{fig3} shows which combinations of model parameters yield
reasonably good matches to the periods observed in GD~358 for this core
composition. The most obvious feature of the parameter-space is the
presence of more than one region that yields a good match to the
observations. Generally, the good fits seem to cluster in two groups
corresponding to thick and thin helium layers. Also obvious are the
parameter-degeneracies in both projections, causing the good fits to fall
along a line in parameter-space rather than on a single point.

The parameter-degeneracy between total mass and fractional helium layer
mass is relatively easy to understand. \cite{bfwh92} showed that the
pulsation periods of trapped modes in white dwarf models are strongly
influenced by the scaled location of the composition transition zone. They
developed an expression showing that these periods are directly
proportional to the {\it fractional} radius of the composition interface. 
As the total mass of a white dwarf increases, the surface area decreases,
so the mass of helium required to keep the interface at the same {\it
fractional} radius also decreases. Thus, a thinner helium layer can
compensate for an overestimate of the mass.

The parameter-degeneracy between mass and temperature is slightly more
complicated. The natural frequency that dominates the determination of
white dwarf model pulsation frequencies is the Brunt-V\"ais\"al\"a
frequency (which reflects the difference between the actual and the
adiabatic density gradients). As the temperature decreases, the matter
becomes more degenerate, so the Brunt-V\"ais\"al\"a frequency in much of
the star tends to zero. The pulsation periods of a white dwarf model in
some sense reflect the average of the Brunt-V\"ais\"al\"a frequency
throughout the star, so a decrease in temperature leads to lower pulsation
frequencies. Higher mass models have higher densities, which generally
lead to higher pulsation frequencies. So an overestimate of the mass can
compensate for the effect of an underestimate of the temperature. 

\begin{figure*}
\epsscale{1.88}
\plotone{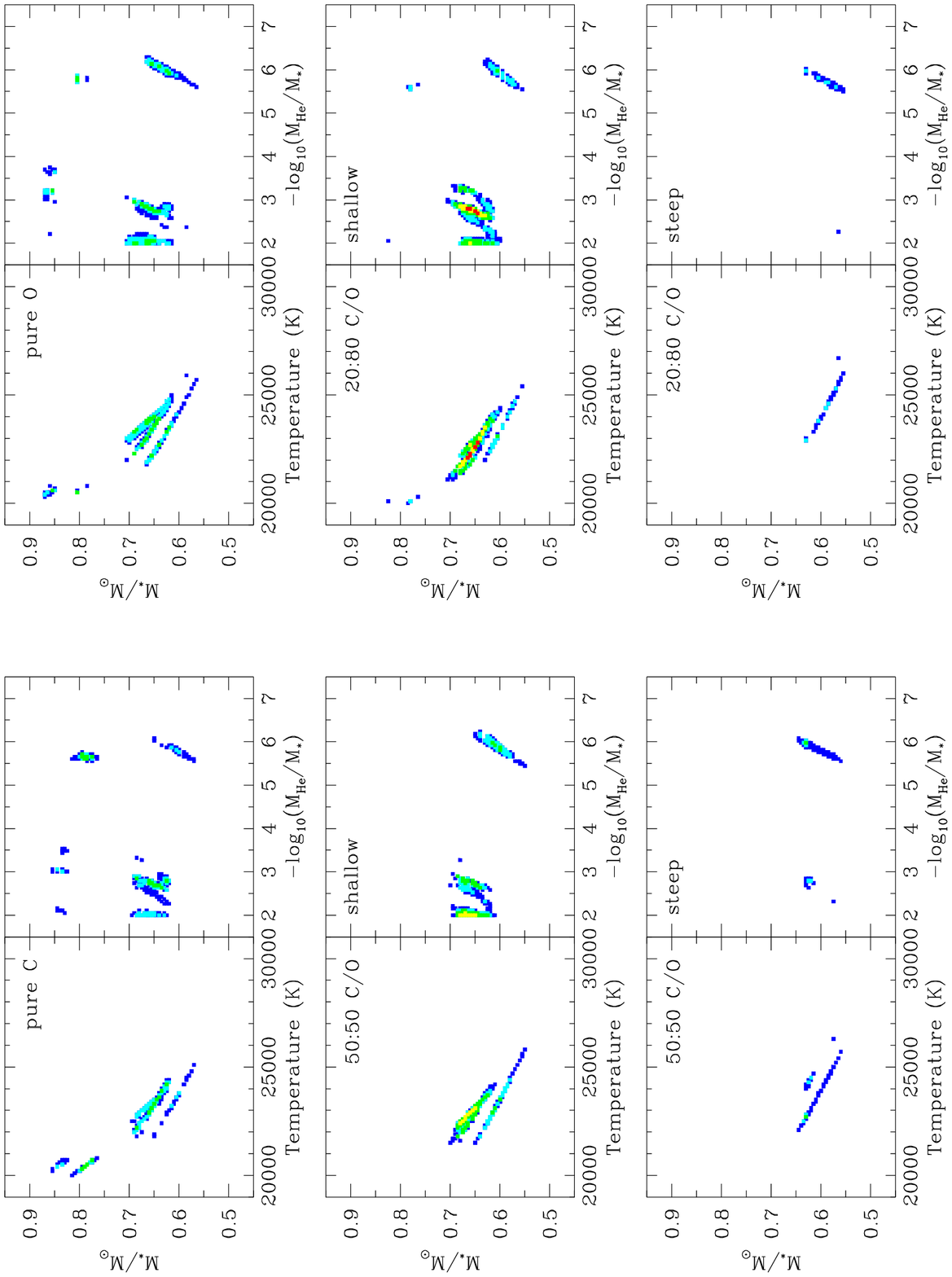}
\figcaption[gd358.fig4.ps]{The families of models found by the GA which 
yield good matches to the periods observed in GD~358 for various core 
compositions and transition profiles. The quality of the fit is indicated 
by the color of the square: ${\rm r.m.s.}<3.0$ seconds (blue), $<2.75$ 
seconds (cyan), $<2.5$ seconds (green), $<2.0$ seconds (yellow), and $<1.65$ 
seconds (red).
\label{fig4}}
\end{figure*}

The results for all six core compositions and transition profiles are
shown in Figure \ref{fig4}, where we have used color to indicate the
absolute quality of each fit. We find that reasonable fits are possible
with every core composition, but excellent fits (indicated by red points
in the figure) are only possible for a specific core composition and
transition profile. Pure C and pure O models appear to have more families
of possible solutions, but the high-mass families have luminosities which
are inconsistent with the observed parallax of GD~358 \citep{har85}. Mixed

\noindent C/O cores generally seem to produce better fits, but transition 
profiles that are steep are much worse than those that are shallow. Among 
the mixed C/O cores with shallow transition profiles, the 20:80 mix produces 
the best fits of all.

The parameters for the best-fit models and measures of their absolute
quality are listed in Table \ref{tab2}. For each core composition, the
best-fit for both the thick and thin helium layer families are shown. As
indicated, several fits can be ruled out based on their luminosities. Our
new best-fit model for GD~358 has a mass and temperature marginally
consistent with those inferred from spectroscopy.

\section{DISCUSSION \& FUTURE WORK}

The genetic algorithm approach to asteroseismology has turned out to be
very fruitful. We are now confident that we can rely on this new approach
to perform global searches and to provide not only objective best-fit
models for the pulsation frequencies of DBV white dwarfs, but also fairly
detailed maps of the parameter-space as a natural byproduct. This approach
can easily be extended to treat the DAV stars and, with a grid of more
detailed starter models, eventually the DOVs. Ongoing all-sky surveys
promise to yield many new pulsating white dwarfs of all classes which will
require follow-up with the Whole Earth Telescope to obtain seismological
data. With the observational and analytical procedures in place, we will
quickly be able to understand the statistical properties of these
ubiquitous and relatively simple objects.

This first application has confirmed that the pulsation frequencies of
white dwarfs really are global oscillations. We have refined our knowledge
of the sensitivity of our models to the structure of the envelope, and we
have confirmed that they are sensitive to the conditions deep in the
interior of the star, as was first demonstrated by the work on
crystallization by \cite{mw99}. 

The fact that our new best-fit solution has a thick helium layer may help
to resolve a long-standing controversy surrounding the evolutionary
connection between the PG~1159 stars and the DBVs. The helium layer mass
for PG~1159-035 from the asteroseismological investigation of \cite{win91}
was $\sim3\times10^{-3}$~\Msun\ while the previous best-fit for GD~358 was
$\sim1.2\times10^{-6}$~\Msun. \cite{dk95} treated this problem by
including time-dependent diffusive processes in their calculations, but
admitted that the presence of the DB gap still remained a problem. The
thick envelope solution could also allow GD~358 to fit comfortably within
an evolutionary scenario leading to a carbon (DQ) white dwarf without
developing an anomalously high photospheric carbon abundance. 

Now that we have seen the significant improvement possible in the fits to
GD~358 by searching globally for various core compositions, we can extend
this method to treat the central C/O ratio as a free parameter. This has
the exciting potential to place meaningful constraints on nuclear reaction
cross-sections. During carbon burning, the triple-$\alpha$ process
competes with the $^{12}{\rm C}(\alpha,\gamma)^{16}{\rm O}$ reaction for
the available $\alpha$-particles. As a consequence, the final ratio of
carbon to oxygen in the core is a measure of the relative cross-sections
of these two reactions \citep{buc96}. With more realistic transition
profiles \citep[e.g.][]{sal97} we hope to use GD~358 to measure this
ratio, which is especially important for understanding type Ia supernovae.

We also plan to test the isotopic separation hypothesis

\begin{table*}
\begin{center}
\tablenum{2}\label{tab2}
\centerline{\sc Table 2}
\centerline{\sc Results for GD 358 Data}\vskip 6pt
\begin{tabular}{lcccl}
\hline\hline
 & \multicolumn{3}{c}{Best-fit Models} & \\ \cline{2-4}
Core Composition & $T_{\rm eff}$ & $M/M_{\odot}$& $\log(M_{\rm He}/M_*)$ & r.m.s.\\
\hline
pure C \dotfill                       & 20,300 & 0.795 & $-$5.66 & 2.17$^a$\\
                                      & 23,100 & 0.655 & $-$2.74 & 2.30 \\
 & & & & \\
50:50 C/O ``shallow''$\ldots$\dotfill & 22,800 & 0.665 & $-$2.00 & 1.76 \\
                                      & 23,100 & 0.610 & $-$5.92 & 2.46 \\
 & & & & \\
50:50 C/O ``steep''\dotfill           & 22,700 & 0.630 & $-$5.97 & 2.42 \\
                                      & 24,300 & 0.625 & $-$2.79 & 2.71 \\
 & & & & \\
20:80 C/O ``shallow''$\ldots$\dotfill & 22,600 & 0.650 & $-$2.74 & 1.50$^b$ \\
                                      & 23,100 & 0.605 & $-$5.97 & 2.48 \\
 & & & & \\
20:80 C/O ``steep'' \dotfill          & 22,900 & 0.630 & $-$5.97 & 2.69 \\
                                      & 27,300 & 0.545 & $-$2.16 & 2.87$^a$ \\
 & & & & \\
pure O \dotfill                       & 20,500 & 0.805 & $-$5.76 & 2.14$^a$ \\
                                      & 23,400 & 0.655 & $-$2.79 & 2.31 \\
\hline\hline
\end{tabular}
\end{center}
\vskip -6pt
\hskip 1.65in{$^a$ Luminosity is inconsistent with observations}

\hskip 1.65in{$^b$ Best-fit solution}
\end{table*}

\noindent in GD~358
\citep{mw00} by investigating whether $^3$He/$^4$He/C/O models provide
significantly better fits than these $^4$He-only models.

We have seen the first indications that our models of white dwarf stars
are incomplete. We hope to identify and correct these weaknesses in the
models by tackling the problem in reverse. Now that we have an objective
global best-fit model, we can investigate what changes to the internal
structure of the model (parameterized by the Brunt-V\"ais\"al\"a
frequency) lead to even better fits than are possible with the current
generation of models. This may lead us to identify regions of the interior
where an improved knowledge of the constitutive physics would be most
useful.

\acknowledgements 
We would like to thank Mike Montgomery, Paul Bradley, S.O. Kepler, and
Craig Wheeler for helpful discussions, Paul Charbonneau for supplying us
with an unreleased version of the PIKAIA genetic algorithm, and Gary
Hansen for arranging the donation of 32 computer processors through AMD. 
This work was supported by grant AST-9876730 from the National Science
Foundation and grant NAG5-9321 from the National Aeronautics \& Space
Administration.

\end{document}